\documentclass[aps,pre,reprintamsfonts,amssymb,amsmath,showpacs,floatfix]{revtex4}

\usepackage[dvips]{graphicx}
\usepackage{mathrsfs}
\usepackage{amsmath,amssymb}

\usepackage{graphicx}
\usepackage{hhline}
\usepackage{dcolumn}% Align table columns on decimal point
\usepackage{bm}% bold math

\usepackage[dvips]{color}

\newcommand{\eref}[1]{Eq.~(\ref{#1})}
\newcommand{\figref}[1]{Fig.~\ref{#1}}

\begin{document}

%\preprint{}

\title{Phase transition without global ordering in a hierarchical scale-free network}
\date{\today}

\author{Takehisa Hasegawa}
\email{hasegawa@m.tohoku.ac.jp}
\affiliation{Graduate School of Information Sciences, Tohoku University, 6-3-09, Aramaki-Aza-Aoba, Sendai, 980-8579, JAPAN.}

\author{Masataka Sato}
\affiliation{Department of Physics, Graduate School of Science,
Hokkaido University, Kita 10-jo Nisi 8-tyome, Sapporo, JAPAN.}

\author{Koji Nemoto}
%\email{nemoto@statphys.sci.hokudai.ac.jp}
\affiliation{Department of Physics, Graduate School of Science,
Hokkaido University, Kita 10-jo Nisi 8-tyome, Sapporo, JAPAN.}

\begin{abstract}
We study the site-bond percolation on a hierarchical scale-free network, 
namely the decorated (2,2)-flower, by using the renormalization group technique.
The phase diagram essentially depends on the fraction of occupied sites.
Surprisingly, when each site is unoccupied even with a small probability, 
the system permits neither the percolating phase nor the nonpercolating phase 
but rather only critical phases.
Although the order parameter always remains zero, 
a transition still exists between the critical phases 
that is characterized by the value of the fractal exponent, 
which measures the degree of criticality; 
the system changes from one critical state to another 
with the jump of the fractal exponent at the transition point.
The phase boundary depends on the fraction of occupied sites. 
When the fraction of unoccupied sites exceeds a certain value, 
the transition line between the critical phases disappears, 
and a unique critical phase remains.
\end{abstract}

%\keywords{}

%\maketitle must follow title, authors, abstract, \pacs, and \words
\maketitle

%\section{introduction}

Hierarchical lattices have occupied a prominent position in the field of statistical physics, 
because the cooperative behaviors on such lattices are exactly solved by the renormalization group (RG) technique.
Recently, hierarchical lattices have also been applied in the context of complex networks 
\cite{albert2002statistical,newman2003structure,boccaletti2006report,dorogovtsev2008critical,barrat2008dynamical}. 
Adjusting the construction rule enables us to generate exactly solvable networks 
having properties common to real networks: 
a scale-free (SF) degree distribution $P(k) \propto k^{-\gamma}$ ($k$ denotes degree), 
a mean shortest path length with logarithmic dependence $l \sim O(\ln N)$ ($N$ being the number of sites), 
and a highly clustering coefficient $C \neq 0$ \cite{rozenfeld2007}.

\begin{figure}[!b]
 \begin{center}
  \includegraphics[width=75mm]{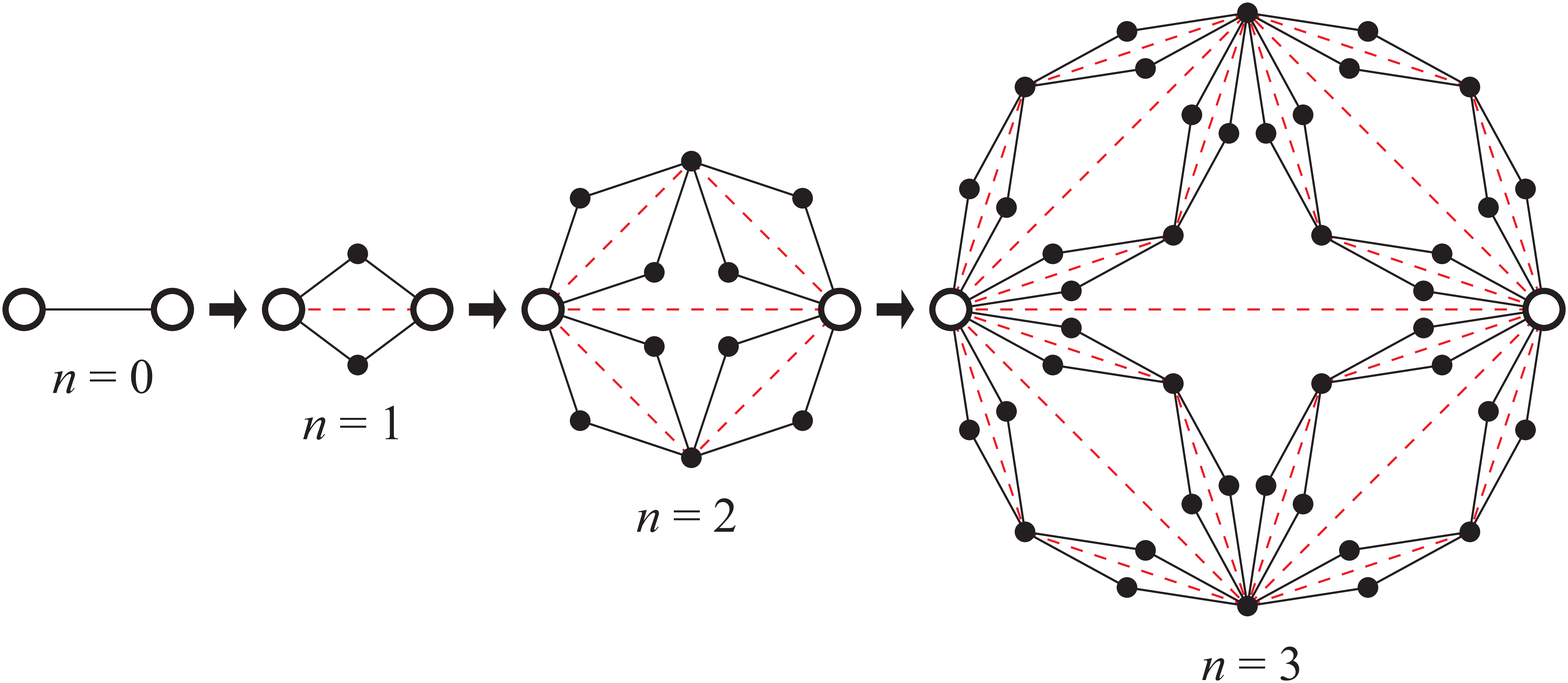}
 \end{center}
 \caption{
(Color online) Realization of the decorated (2,2)-flower $\tilde{F}_n$ (and $F_n$) 
with $n=0, 1, 2, 3$. 
The root sites are indicated by the open circles. 
$F_n$ (solid black lines) is recursively constructed (see the text), and 
$\tilde{F}_n$ is obtained by adding the shortcuts (red dashed lines) to $F_n$. 
Note that the shortcuts remain the same in each iteration.
}
 \label{rule}
\end{figure}

Dynamics (such as percolations and spin systems) on hierarchical SF networks 
provide an impetus to further investigate the relation between network topology and dynamics 
because of the abnormal phase transitions on such networks. 
Hinczewski and Berker \cite{hinczewski2006inverted} analyzed the Ising model on a hierarchical SF network, 
called the decorated (2,2)-flower (\figref{rule}), to show that the system undergoes 
an inverted Berezinskii-Kosterlitz-Thouless (BKT) transition, 
which means that the system shows 
a BKT singularity {\it above} the transition temperature.
A similar abnormal behavior holds for the case of bond percolation \cite{hasegawa2010generating,berker2009critical}. 
The bond percolation on the decorated (2,2)-flower has the percolating phase, 
which contains a unique giant component whose size is of the order $O(N)$, 
and the \textit{critical phase} (also known as the
partially ordered phase or the ``patchy'' phase for the Hanoi network \cite{boettcher2009patchy}). 
In terms of RG, the critical phase is characterized by the RG flow 
converging onto the line of the nontrivial stable fixed point 
in contrast to the nonpercolating phase containing only finite size clusters.

%%%%%%%%%%%%%%%%%%%%%%%%%%%%%%%%%%%%%%
%%%%%%%%%%%%%%%%%%%%%%%%%%%%%%%%%%%%%%

In the critical phase, where the order parameter is zero, the system is characterized by the fractal exponent $\psi$ 
\cite{hasegawa2010generating, NAG}.
The fractal exponent $\psi$ is defined as 
the logarithmic derivative of the mean size $S_{\rm max}(N)$ 
of the largest cluster, 
$\psi \equiv \lim_{N \to \infty} \log_N S_{\rm max}(N)$,
which mimics $d_f/d$ for $d$-dimensional Euclidean lattice systems, 
where $d_f$ is the fractal dimension of the largest clusters. 
In the critical phase $\psi$ takes a positive value less than one, $0<\psi<1$, 
while $\psi=1$ indicates the percolating phase and $\psi=0$ the nonpercolating phase.
The mean number $n_s$ of clusters with size $s$ per site (or the cluster size distribution, in short) obeys the power-law 
$n_s \propto s^{-\tau}$ with varying exponent $\tau=1+\psi^{-1}$ 
over the entire region of the critical phase.
Such a critical phase is also observed in growing random networks 
(see \cite{hasegawa2010scaling} and references therein).

%%%%%%%%%%%%%%%%%%%%%%%%%%%%%%%%%%%%%%
%%%%%%%%%%%%%%%%%%%%%%%%%%%%%%%%%%%%%%

\begin{figure*}[!tbp]
 \begin{center}
  \includegraphics[width=60mm]{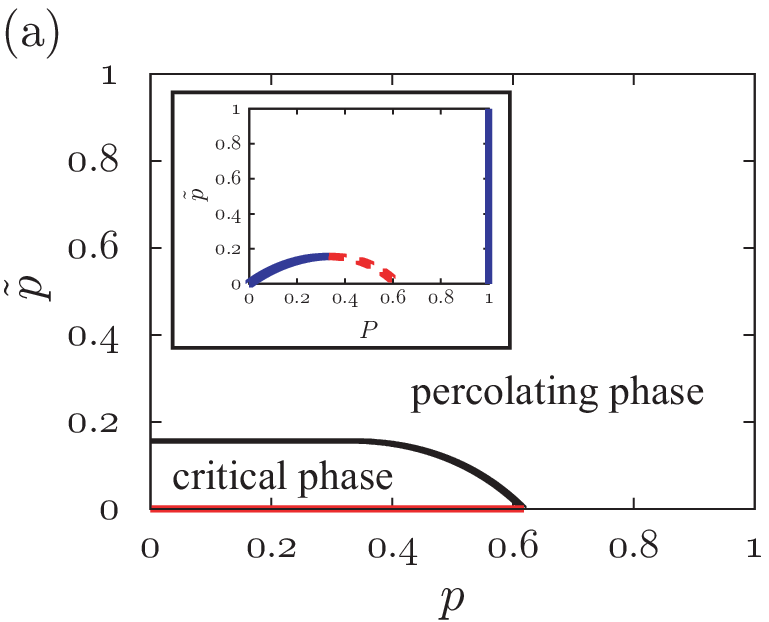}
  \includegraphics[width=60mm]{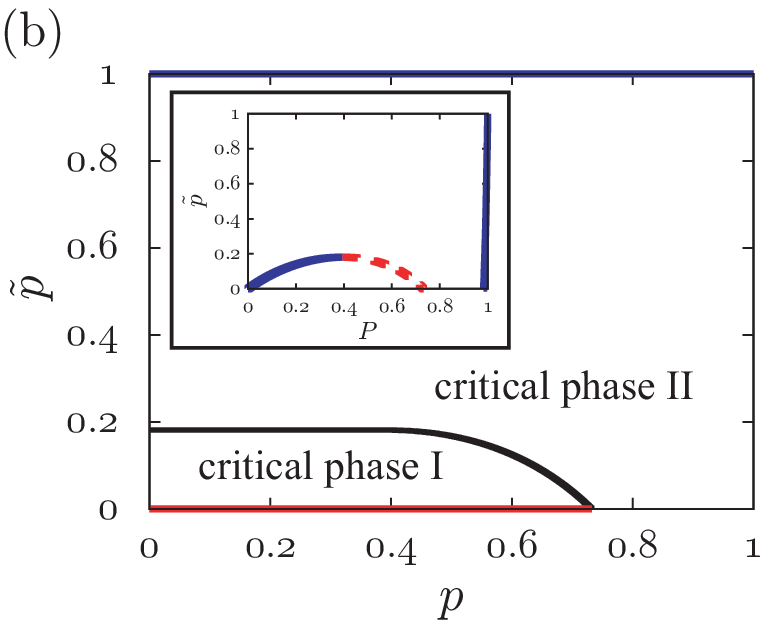}
  \includegraphics[width=60mm]{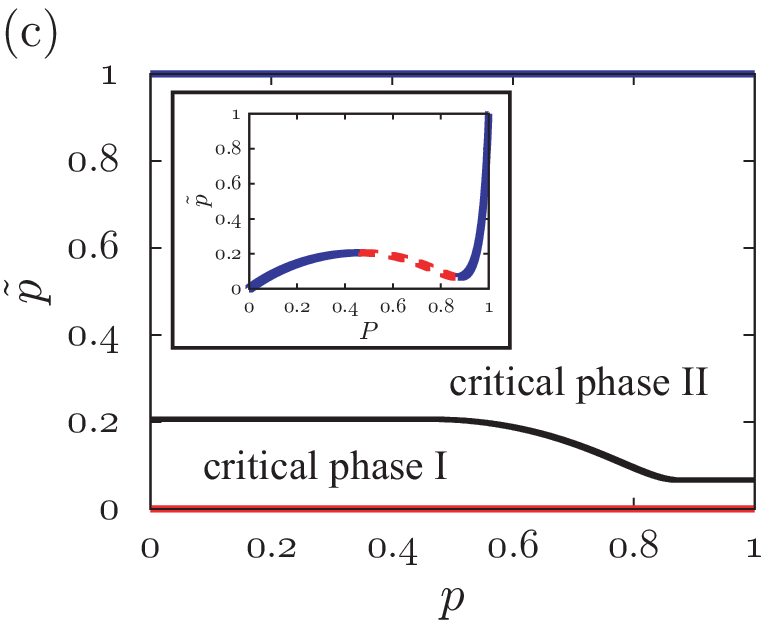}
  \includegraphics[width=60mm]{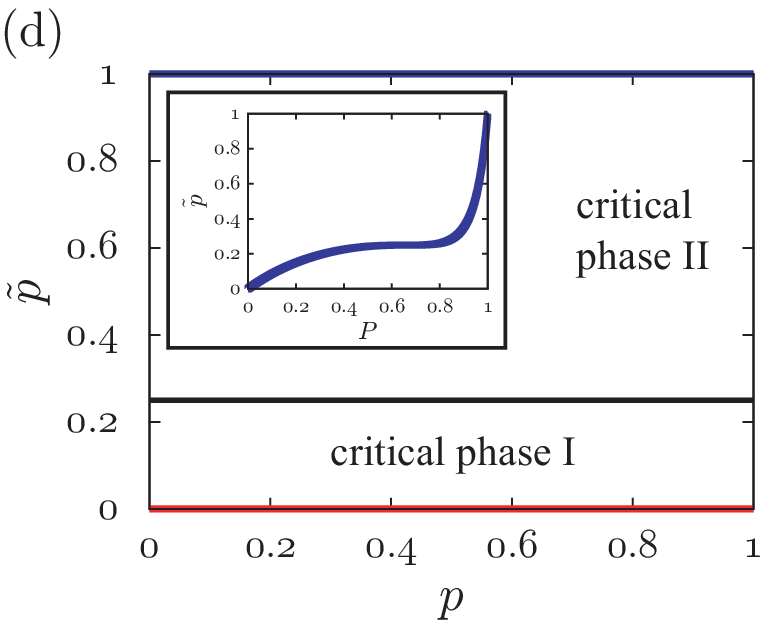}
 \end{center}
 \caption{
(Color online) Phase diagrams for several values of $p_s$: 
(a) $p_s=1.0$, (b) $p_s=0.9$, (c) $p_s=53/64$, and (d) $p_s=3/4$.
The bold black line indicates the phase boundary.
The blue line at $\tilde{p}=1$ indicates the percolating phase, 
and the red line at $\tilde{p}=0$ indicates the nonpercolating phase. 
Each inset shows the lines of stable and unstable fixed points.
The solid blue and red dashed lines 
indicate $p^*(\tilde{p},p_s)$ and $p^{**}(\tilde{p},p_s)$ (stable fixed point) 
and $p_c(\tilde{p},p_s)$ (unstable fixed point), respectively.
}
 \label{fig:Phase}
\end{figure*}

%%%%%%%%%%%%%%%%%%%%%%%%%%%%%%%%%%%%%%
%%%%%%%%%%%%%%%%%%%%%%%%%%%%%%%%%%%%%%

Previous studies revealed that the phase diagram of dynamics on 
complex networks is not simply described by globally ordered state and 
disordered state. The concept of partially ordering may be needed for 
the complete understanding of the dynamics on the complex networks.
In this paper, we investigate the site-bond percolation on the decorated 
(2,2)-flower to demonstrate the existence of a hidden phase transition 
without globally ordered state.
In this network, 
defects of sites may have significantly different effects from those of the defects of bonds, 
because the defect of a site with a high degree is directly connected to the removal of infinitely many bonds, 
and also, the sites with high degrees are interconnected. 
We demonstrate that the site dilution dramatically changes the phase diagram.
Surprisingly, even if the fraction of the removed sites is very small, 
global ordering is broken, 
and only the partially ordered state is permitted.
Furthermore, although the order parameter is zero in the entire region, 
a transition between two different critical phases still exists, 
which is characterized by a discontinuous jump of the fractal exponent.
When the fraction of the removed sites exceeds a certain value, 
the boundary between the two critical phases disappears, and a unique critical phase remains.

%\section{model}

%%%%%%%%%%%%%%%%%%%%%%%%%%%%%%%%%%%%%%
%%%%%%%%%%%%%%%%%%%%%%%%%%%%%%%%%%%%%%

\begin{figure}[!tbp]
 \begin{center}
  \includegraphics[width=60mm]{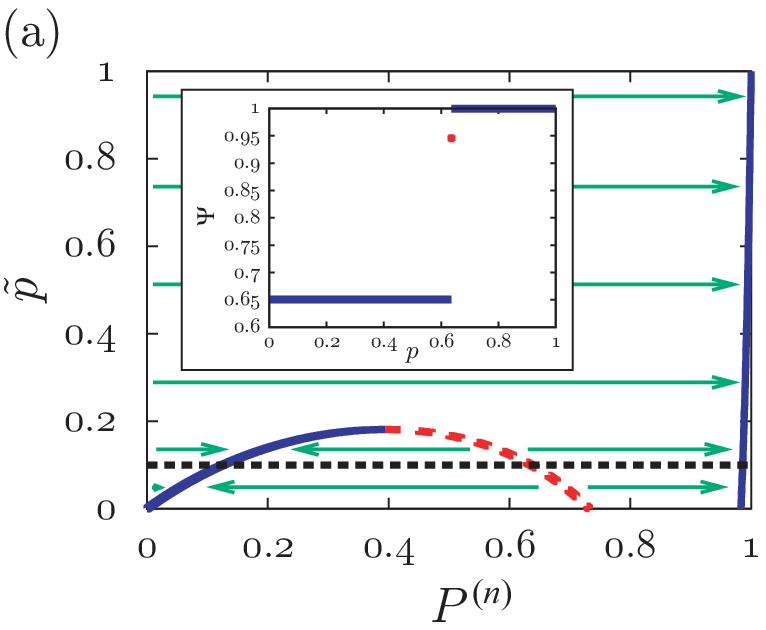}
  \includegraphics[width=50mm]{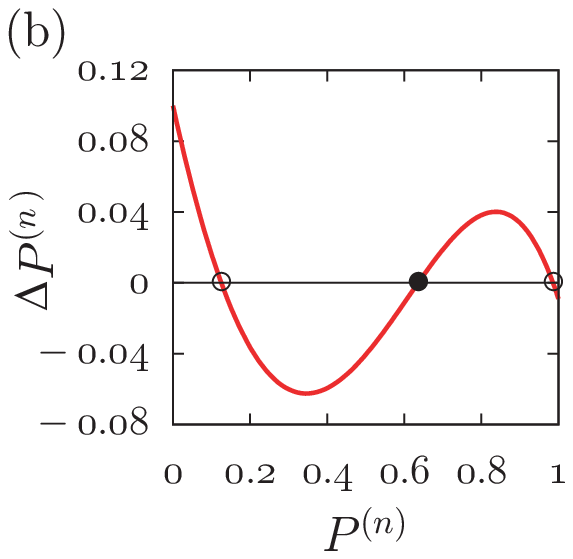}
 \end{center}
 \caption{
(Color online) (a) Flow diagram for $p_s=0.9$. 
Arrows indicate the flow of \eref{sP^{(n)}+1}. 
The solid blue and red dashed lines indicate the stable and unstable fixed points, respectively.
The inset shows the fractal exponent $\psi$ as a function of $p$.
Here $\psi=0.6504, 0.9457$, and $0.9994$
for $p<p_c(\tilde{p},p_s), p=p_c(\tilde{p},p_s)$, and $p>p_c(\tilde{p},p_s)$, respectively.
(b) $\Delta P^{(n)}$ as a function of $P^{(n)}$.
The open and filled circles indicate the unstable and stable fixed points, respectively.
%(c) Cluster size distribution $n_s$ at $p=p^*(\tilde{p},p_s)$ (blue triangles), $p=p_c(\tilde{p},p_s)$ (green squares), and $p=p^{**}(\tilde{p},p_s)$ (red circles).
%In the inset of (a) and (b), and (c), 
In the inset of (a) and also in (b), 
we set $p_s=0.9$ and $\tilde{p}=0.1$ 
[the horizontal dotted line in the main panel of (a)].
}
 \label{fig:Psi}
\end{figure}
%

%%%%%%%%%%%%%%%%%%%%%%%%%%%%%%%%%%%%%%
%%%%%%%%%%%%%%%%%%%%%%%%%%%%%%%%%%%%%%

A special class of hierarchical SF networks, the (decorated) ($u,v$)-flower, was introduced in \cite{rozenfeld2007}. 
The (2,2)-flower $F_n$ of the $n$-th generation is recursively constructed as follows (Fig.~\ref{rule}): 
At $n=0$, $F_0$ consists of two sites connected by a bond. 
We call these sites \textit{root sites}.
For $n \ge 1$, $F_n$ is obtained from $F_{n-1}$, 
such that each existing bond in $F_{n-1}$ is replaced by two parallel paths consisting of two bonds each. 
As illustrated in Fig.~\ref{rule}, the decorated (2,2)-flower $\tilde{F}_n$ of the $n$th generation 
is given by adding the shortcuts (red dashed lines) to $F_n$. 
The number of sites $N_n$ of $\tilde{F}_n$ is $N_n=2(4^n+2)/3$, 
and $\tilde{F}_n$ has $P(k) \propto k^{-3}$, $\ell \propto \ln N$, and $C \sim 0.820$ \cite{rozenfeld2007}.

Let us consider the site-bond percolation on the decorated (2,2)-flower.
Each site is occupied with the probability $p_s$.
The open-bond probability of bonds constructing $F_n$ is $p$, 
and that of the shortcuts is $\tilde{p}$.
The phase diagram of the model is obtained by the RG technique 
\cite{rozenfeld2007,hasegawa2010generating,berker2009critical}.
Let $P^{(n)}$ be the probability that both roots are 
in the same cluster after a site-bond percolation trial on $\tilde{F}_n$.
Here we do not inquire whether the two roots are occupied or not.
In the large size limit, the system is regarded as \textit{percolating} 
if the percolation probability $P = \lim_{n\to\infty} P^{(n)}$ is nonzero \cite{rozenfeld2007}.
We find the recursive equation of $P^{(n)}$ as 
\begin{eqnarray}
 P^{(n+1)}=\tilde{p}+\tilde{q} \left[2p_{s}(P^{(n)})^2-p_{s}^2(P^{(n)})^4\right],  \label{sP^{(n)}+1}
\end{eqnarray}
where $\tilde{q} \equiv 1-\tilde{p}$, 
and the initial value is set to $P^{(0)}=p$.
The fixed point $P=P^{(n)}=P^{(n+1)}$ of \eref{sP^{(n)}+1} satisfies 
\begin{eqnarray}
 \tilde{p}(P)=1-\frac{1-P}{(1-p_{s}P^2)^2}. \label{site_fix}
\end{eqnarray}
We obtain the phase diagram from 
the RG technique [\eref{sP^{(n)}+1}] 
and the fixed-point condition [Eq.~(\ref{site_fix})].

%\section{result}
The phase diagram for $p_s=1$, i.e., for the bond percolation with $p$ and $\tilde{p}$, is already known 
\cite{hasegawa2010generating,berker2009critical} 
[Fig.~\ref{fig:Phase}(a)].
%The system with $p_s=1$ has the critical phase and the percolating phase. 
For a fixed $\tilde{p}$ ($0<\tilde{p}<\tilde{p}_c=5/32$), 
there are two stable fixed points: $0<P=p^*(\tilde p,p_s)<1$ and $P=1$,
which correspond to the critical and percolating phases, respectively. 
Furthermore, one unstable fixed point between these two 
fixed points gives the phase boundary: $P=p_c(\tilde p,p_s)$.
For $\tilde p >  \tilde p_c$, there is only one stable fixed point at $P=1$, 
so that the system is always percolating.

When a site dilution is added ($3/4 < p_s <1$), the phase diagram dramatically changes [\figref{fig:Phase}(b)].
The flow diagram with $p_s=0.9$ is shown in \figref{fig:Psi}(a).
For a given $\tilde{p}$ ($0 < \tilde{p} <\tilde{p}_c$), 
there are two nontrivial stable fixed points at $P=p^*(\tilde{p},p_s), p^{**}(\tilde{p},p_s)$ 
and one unstable fixed point at $P=p_c(\tilde{p},p_s)$, 
where $0<p^*(\tilde{p},p_s)<p_c(\tilde{p},p_s)<p^{**}(\tilde{p},p_s)<1$. 
Here the (in)stabilities of the fixed points are given by 
$\Delta P^{(n)} \equiv P^{(n+1)}-P^{(n)}$ [\figref{fig:Psi}(b)]. 
We find that all RG flows converge onto nontrivial fixed points ($0<P<1$). 
The system has neither the percolating phase nor the nonpercolating phase, 
but rather, it has two different critical phases, 
which we call critical phases I and II.
In the entire region, $n_s$ obeys a power-law 
with a corresponding exponent (not shown).
%(\figref{fig:Psi}(c)). 
Also, the mean cluster size $\langle s \rangle= \sum_s s^2 n_s$ always diverges, because $\psi>1/2$, 
which means $\tau <3$ (see below).
For $\tilde{p}>\tilde{p}_c$, one stable fixed point $P=p^{**}(\tilde{p},p_s)$ exists, 
and the system is always in the same critical phase.

%%%%%%%%%%%%%%%%%%%%%%%%%%%%%%%%%%%%%%
%%%%%%%%%%%%%%%%%%%%%%%%%%%%%%%%%%%%%%

\begin{figure}[htb]
 \begin{center}
 \vspace{0.1cm}
  \includegraphics[width=60mm]{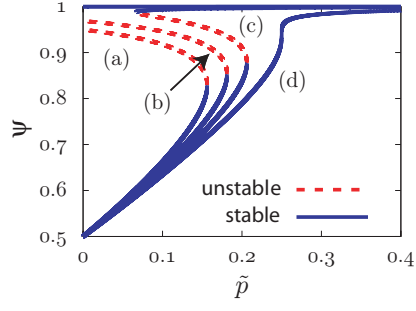}
 \end{center}
 \caption{
Fractal exponent $\psi$ as a function of $\tilde{p}$ for 
several values of $p_s$: 
(a) $p_s=1.0$, (b) $p_s=0.9$, (c) $p_s=53/64$, and (d) $p_s=3/4$.
}
 \label{fig:PsiSeveral}
\end{figure}
%

%%%%%%%%%%%%%%%%%%%%%%%%%%%%%%%%%%%%%%
%%%%%%%%%%%%%%%%%%%%%%%%%%%%%%%%%%%%%%

A kind of phase transition between critical phases may exist when we pass an unstable fixed point, 
even though the order parameter remains zero in the entire parameter region.
Let us calculate the fractal exponent $\psi$ of the mean size $S_{\rm root}(N)$ 
of the cluster to which a root belongs, defined as 
$\psi = \lim_{n \to \infty} {\rm d}\ln S_{\rm root}(N_n)/{\rm d}\ln N_n$. 
We consider three quantities on $F_n$: 
(1) the probability $t_k^{(n)}(p)$ that 
both roots are connected to the same cluster of size $k$,
(2) the probability $s_{k,l}^{(n)}(p)$ that the left (right) root is connected to a cluster of size $k$ ($l$) 
but that these clusters are not the same, 
and 
(3) the mean number $u_k^{(n)}(p)$ of clusters of size $k$ to which neither of the roots is connected.
For the sake of convenience, the roots are not counted 
in the cluster size $k$ or $l$ for $t_k^{(n)}(p)$ and $s_{k,l}^{(n)}(p)$. 
We introduce the generating functions 
$T_n(x)$, $S_n(x,y)$, and $U_n(x)$ 
for $t_k^{(n)}(p)$, $s_{k,l}^{(n)}(p)$, and $u_k^{(n)}(p)$, 
where $T_n(x) \equiv \sum_{k=0}^{\infty}t_k^{(n)}(p)x^k$,
$S_n(x,y)\equiv \sum_{k=0}^{\infty}\sum_{l=0}^{\infty}s_{k,l}^{(n)}(p)x^ky^l$, 
and $U_n(x) \equiv \sum_{k=0}^{\infty}u_k^{(n)}(p)x^k$. 
The self-similar structure of $F_n$ allows us to obtain recursion relations for these generating functions: 
\begin{widetext}
\begin{eqnarray}
T_{n+1}(x)&=&
(p_s^2x^2+2p_sq_sx)T_n^4(x)+4p_s^2x^2T_n^3(x)S_n(x,x) 
+4p_sq_sxT_n^3(x)S_n(1,x)
+2p_sxT_n^2(x)S_n(x,1)S_n(1,x), 
  \label{T_n+1_s}
\\
 S_{n+1}(x,y)&=& 
\left\{S_n(x,1)S_n(1,y)+p_sS_n(x,y)\left[xT_n(x)+yT_n(y)\right]
+q_s\left[T_n(x)T_n(y)+T_n(x)S_n(1,y)+T_n(y)S_n(x,1)\right]\right\}^2,
 \label{S_n+1_s} 
\\
 U_{n+1}(x)&=&4U_n(x)+2\left\{p_sxS_n(1,x)S_n(x,1)+q_s\left[S_n(1,x)+S_n(x,1)\right]\right\},
 \label{U_n+1_s}
\end{eqnarray}
\end{widetext}
where $q_s \equiv 1-p_s$.
The corresponding generating functions 
$\tilde{T}_n(x)$, $\tilde{S}_n(x,y)$, and $\tilde{U}_n(x)$ on $\tilde{F}_n$ are given by 
$\tilde{T}_n(x)=T_n(x)+\tilde{p} S_n(x,x)$, 
$\tilde{S}_n(x,y)=\tilde{q} S_n(x,y)$, and $\tilde{U}_n(x)=U_n(x)$, respectively.
Here we denote the mean fractions of the clusters 
including two roots and either of the roots by 
$\tau_n=N_n^{-1}\frac{\rm d}{{\rm d}x}\tilde{T}_n(x)|_{x=1}$ and 
$\sigma_n=N_n^{-1}\frac{\rm d}{{\rm d}x}\tilde{S}_n(x,x)|_{x=1}$, respectively. 
Similarly to \cite{hasegawa2010generating}, 
we obtain the recursive equation for $\tau_n$ and $\sigma_n$ as 
\begin{widetext}
\begin{eqnarray}
  \begin{pmatrix}
   \sigma_{n+1}  \\
   \tau_{n+1} 
  \end{pmatrix}
  &=& \frac{{N}_n}{{N}_{n+1}} 
  \begin{pmatrix}
    2\tilde{q}(1-p_s(P^{(n)})^2)(1+p_sP^{(n)})  & 
    4\tilde{q}(1-p_s(P^{(n)})^2)(1-p_sP^{(n)})  \\
    2(1+p_sP^{(n)})\left[1-\tilde{q}(1-p_s(P^{(n)})^2)\right]  &
    4\left[1-\tilde{q}(1-p_s(P^{(n)})^2)(1-p_sP^{(n)})\right]  
  \end{pmatrix}
  \begin{pmatrix}
    \sigma_{n}  \\
    \tau_{n} 
  \end{pmatrix} \nonumber \\
  &&+\frac{1}{{N}_{n+1}}
  \begin{pmatrix}
    4\tilde{q}p_sP^{(n)}Q^{(n)}(1-p_s(P^{(n)})^2)  \\
    2p_sP^{(n)}\left[2-P^{(n)}-2\tilde{q}Q^{(n)}(1-p_s(P^{(n)})^2)\right] 
  \end{pmatrix} \\
  &\sim &
   \begin{pmatrix}
   \frac{1}{2}\frac{Q(1+p_sP)}{1-p_sP^2} &
\frac{Q(1-p_sP)}{1-p_sP^2} \\
   \frac{1}{2}\frac{P(1+p_sP)(1-p_sP)}{1-p_sP^2} &1-\frac{Q(1-p_sP)}{1-p_sP^2}  
  \end{pmatrix}
  \begin{pmatrix}
   \sigma_{n}  \\
   \tau_{n}
  \end{pmatrix}
, \quad {\rm for \,\,} n \gg 1, \label{e15}
\end{eqnarray}
\end{widetext}
where $Q \equiv 1-P$ and $Q^{(n)} \equiv 1-P^{(n)}$, and we used the fixed-point condition [Eq.~(\ref{site_fix})].
Then, the largest eigenvalue $\lambda$ of \eref{e15} gives 
the fractal exponent as $\psi =1+ \ln \lambda/ \ln 4$.
The inset of Fig.~\ref{fig:Psi} (a) shows 
the fractal exponent $\psi$ for $p_s=0.9$ and $\tilde{p}=0.1$.
When $p$ increases, 
the fractal exponent jumps from a nonzero value 
$\psi=0.6504$ 
to another larger value 
$\psi=0.9994$ 
at the transition point $p_c(\tilde{p},p_s)$.

%%%%%%%%%%%%%%%%%%%%%%%%%%%%%%%%%%%%%%
%%%%%%%%%%%%%%%%%%%%%%%%%%%%%%%%%%%%%%

\begin{figure}[!tp]
 \begin{center}
  \includegraphics[width=60mm]{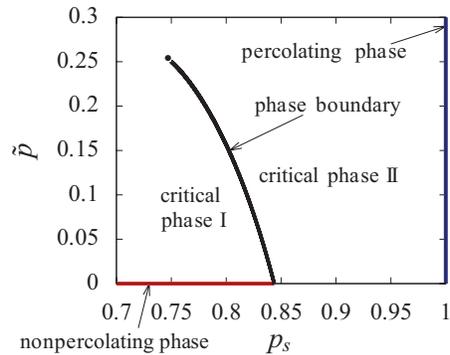}
 \end{center}
 \caption{
(Color online) Phase diagram of the case with $p=1$.
The bold black line indicates the transition line between the critical phases.
The blue line (at $p_s=1$) and the red line (at $\tilde{p}=0$) 
indicate the percolating and nonpercolating phases, respectively.
}
 \label{fig:p1fix}
\end{figure}

%%%%%%%%%%%%%%%%%%%%%%%%%%%%%%%%%%%%%%
%%%%%%%%%%%%%%%%%%%%%%%%%%%%%%%%%%%%%%

The phase diagram and the corresponding $\psi$-curve depends on $p_s$ 
(Figs.~\ref{fig:Phase} and \ref{fig:PsiSeveral}).
When $p_s$ decreases to below $p_s=3/4$, 
the unstable fixed point disappears, 
and the two lines of the stable fixed points merge.
Then $\psi$ increases monotonically with $\tilde{p}$, but never jumps (\figref{fig:PsiSeveral}).
For $p_s<3/4$, the boundary between critical phases I and II disappears, 
and the unique critical phase remains.

Finally, we mention the phase diagram for $p=1$ (\figref{fig:p1fix}).
Figure \ref{fig:p1fix} also indicates that the percolating phase is destroyed by infinitesimal site dilution. 
This destruction is related to the fact that the sites with high degrees are interconnected in a hierarchical manner.
Moreover, the phase boundary between critical phases I and II disappears at $(\tilde{p},p_s)=(1/4, 3/4)$.
This resembles the phase diagram of a vapor-liquid transition.

To summarize, we have investigated the site-bond percolation on the decorated (2,2)-flower. 
The phase diagram essentially depends on the fraction of the occupied sites $p_s$: 
the system has the percolating and critical phases at $p_s=1$, 
two critical phases (I, II) for $3/4< p_s <1$, 
and a unique critical phase for $p_s <3/4$.
We suggest that critical phase I is caused by shortcut insertion, 
while critical phase II is due to the site dilution.
The shortcuts of the decorated (2,2)-flower, 
which connect high-degree nodes, 
easily give rise to a partially ordered state. 
Further, site dilutions in such connections immediately break a globally ordered state into a partially ordered state.

We demonstrated the existence of a new type of phase transition without global ordering, namely, a transition between two critical phases. 
Our result seems to hold for the other decorated ($u,v$)-flower ($u,v>1$), 
although it is an open question whether such a transition exists on other complex network models.
Site dilutions in complex networks may also have extreme effects 
on other dynamics (e.g., spin systems and the contact process).
We hope that further studies on phase transition without global ordering 
give a new perspective on the critical phenomena emerging in complex networks.

\end{document}